%% file: eprint.tex
\documentclass[12pt]{article}
\usepackage{graphicx}

\newcommand\pubnumber{}
\newcommand\pubdate{}

\def\institute{Universit\'e catholique de Louvain, Louvain-la-Neuve, BELGIUM}

\def\Title#1{\begin{center} {\Large #1 } \end{center}}
\def\Author#1{\begin{center}{ \sc #1} \end{center}}
\def\Address#1{\begin{center}{ \it #1} \end{center}}

\newcommand\pubblock{\rightline{\begin{tabular}{l} \pubnumber\\
         \pubdate  \end{tabular}}}
\newenvironment{Abstract}{\begin{quotation}  }{\end{quotation}}
\newenvironment{Presented}{\begin{quotation} \begin{center} 
             PRESENTED AT\end{center}\bigskip 
      \begin{center}\begin{large}}{\end{large}\end{center} \end{quotation}}

\input econfmacros.tex

\begin{document}
\begin{titlepage}
\pubblock

\vfill
\Title{
Observation of top quark production \\ in proton-nucleus collisions
}
\vfill
\Author{ Georgios Konstantinos Krintiras \\ On behalf of the CMS Collaboration}
\Address{\institute}
\vfill
\begin{Abstract}
The multi-TeV energies available at LHC have opened up the possibility to measure, for the first time, various large-mass elementary particles in nuclear collisions.
The current study presents the first observation of top quark--the heaviest elementary particle in the standard model--using proton-lead collisions.
The measurement is based on a data set whose integrated luminosity amounts to 174~nb$^{-1}$, as recorded by CMS at a center-of-mass energy per nucleon pair of 8.16 TeV. 
The pair production process is measured using events with exactly one isolated lepton, electron or muon, and at least four jets, leading to a cross section of $45\pm8\ \rm{nb}$. 
This is well compatible with theoretical predictions from perturbative quantum chromodynamics at next-to-next-to-leading order with soft gluon resummation at next-to-next-to-leading logarithmic accuracy. 
The statistical significance of the signal against the background-only hypothesis is above five standard deviations.
\end{Abstract}
\vfill
\begin{Presented}
$10^{th}$ International Workshop on Top Quark Physics\\
Braga, Portugal,  September 17--22, 2017
\end{Presented}
\vfill
\end{titlepage}
\def\thefootnote{\fnsymbol{footnote}}
\setcounter{footnote}{0}

\section{Introduction}
The top quark, the heaviest elementary particle in the standard model with mass ($\rm{m_{t}}$) as much as a tungsten atom, 
has been subjected to detailed scrutiny using hadron-hadron collisions. 
Until recently, top quark measurements remained out of reach in nuclear collisions due to the reduced amount
of integrated luminosity produced during the first period at the LHC, and the relatively low nucleon-nucleon center-of-mass energies ($\sqrt{s_{\rm{NN}}}$)
available at the BNL RHIC. 
Novel studies of top quark cross sections have finally become feasible with the 2016 LHC proton-lead (pPb) run at $\sqrt{s_{\rm{NN}}}=8.16\ \rm{TeV}$,
surpassing in performance almost eight times the designed instantaneous luminosity and delivering to CMS~\cite{CMS} 174~nb$^{-1}$ of pPb collision data.   

At hadron colliders, top quarks are produced either in pairs ($\rm{t}\bar{\rm{t}}$), predominantly through the strong, or singly, through the weak interaction.
As such, their production cross sections are computable with great accuracy in perturbative quantum chromodynamics (QCD), 
rendering them theoretically precise probes of the gluon density, both free and bound, in an unexplored kinematic regime around Bjorken-$x \approx 2\rm{m_{t}}/\sqrt{s_{\rm{NN}}}$
and high virtualities $Q^2 \approx \rm{m_{t}}^2 $. The production of top quarks therefore provides information on the nuclear parton distribution functions (nPDF) that is complementary to that
obtained  through  studies  of  dijet and electroweak  boson  production. 
In nuclear collisions, the rate of top quark interactions with the color fields stretched among the involved partons are enhanced, when compared to hadron-hadron collisions.
The reconstruction of $\rm{m_{t}}$ in high-density regimes will thus provide interesting insights also in non-perturbative QCD effects, i.e., the color flow 
between the t and $\bar{\rm{t}}$ quarks, their decay products, and the underlying event from multi-parton interactions and beam remnants~\cite{david}.

Once produced, the top quark decays promptly without hadronizing into a W boson plus a bottom quark.
When one W boson decays leptonically ($\ell=\mu,\rm{e}$) and the other hadronically ($\mathrm{j} \mathrm{j}^{\prime}$), the emerging $\ell$+jets final state 
presents a typical signature of one isolated charged lepton and momentum imbalance from the unobserved neutrino, on the one hand, 
two light-quark jets, on the other hand, and two b (``tagged'') jets from the initial decay. 
The $\rm{t}\bar{\rm{t}}$ cross section can be then extracted from a combined fit of the invariant mass of the $\mathrm{j} \mathrm{j}^{\prime}$ system. 
The correlation in phase space of the light-quark jets carries a distinctive hallmark with respect to
the main backgrounds that are controlled in different categories of events with zero, one, or at least two b-tagged jets.

\section{Proton-lead collisions at LHC}

\begin{figure}[htb]
\centering
\includegraphics[scale=0.25]{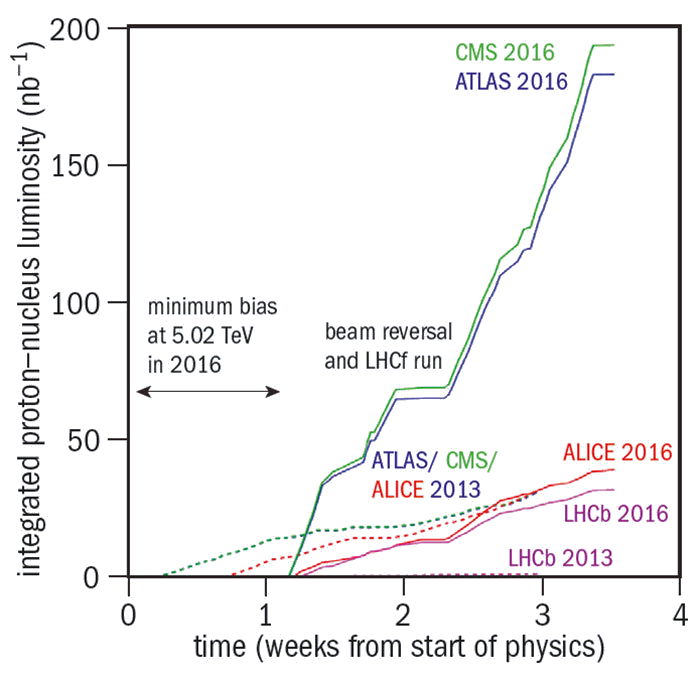}
\includegraphics[scale=0.215]{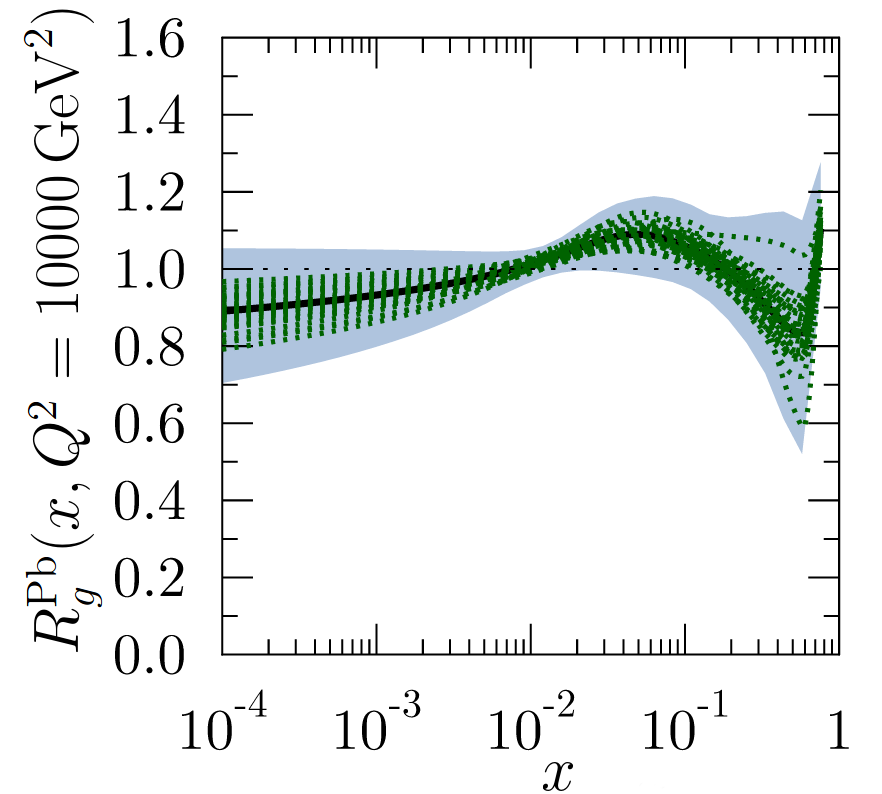}
\caption{
(left) Accumulation of integrated luminosity in each LHC experiment in the pPb runs for years 2013 and 2016~\cite{Jowett:IPAC2017}.
(right) The EPPS16 nuclear modifications for the bound gluon PDF in Pb nucleus at the parametrization scale $Q^2=10^4$~$\rm{GeV^2}$.
The thick black curves correspond to the central fit values, the dotted curves to the individual error sets, and 
the total uncertainties are shown as blue bands~\cite{Eskola}.}
\label{fig:pPb}
\end{figure}

Asymmetric collisions of $^{208}\rm{Pb}^{82+}$ nuclei with protons had not been included in the initial LHC design.
However, unexpected discoveries in small collision systems, reminiscent of flow-like collective phenomena, engaged further investigations~\cite{Salgado}.
After the short, yet remarkable, pilot physics run in 2012, and the first full one-month run in early 2013, 
the second full pPb run took place in late 2016, in order to deliver data at $\sqrt{s_{\rm{NN}}}=8.16\ \rm{TeV}$ for each direction of the beams. 
Apart from complex bunch filling schemes due to the generation of the beams from two separate injection paths, 
the distinct feature of operation with asymmetric collisions at LHC is the difference in revolution and cavity frequencies.
Given that the colliding bunches have significantly different size and charge, both beams are displaced transversely, 
onto opposite-sign off-momentum orbits, and longitudinally, to restore collisions at the proper interaction points, 
during a process colloquially known as ``cogging''~\cite{Jowett:IPAC2017}. 
The long-term integrated luminosity goal of 100~nb$^{-1}$ has been clearly surpassed (Fig.~\ref{fig:pPb}, left), 
rendering the 2016 pPb run the baseline for several years.

The binding energies of nucleons in the nucleus are several orders of magnitude smaller than the momentum transfers of deep-inelastic 
scattering (DIS). Therefore, such a ratio naively should be close to unity, except for small corrections for the Fermi motion of nucleons in the nucleus. 
Contrary to expectations, the EMC experiment discovered a declining slope to the ratio~\cite{EMC}, a result later confirmed with high-precision 
electron- and muon-scattering DIS data. 
The conclusions from the combined experimental evidence demonstrates the consistency in modifying the free nucleon PDFs, $f_{i}^{\rm{p}}(x,Q^2)$, at low $Q^2$ and 
letting the Dokshitzer-Gribov-Lipatov-Altarelli-Parisi evolution to take care of the momentum transfer dependence, i.e., 
$f_{i}^{\rm{p/A}}(x,Q^2)=R_{i}^{A}(x,Q^2)f_{i}^{\rm{p}}(x,Q^2)$, where $R_{i}^{A}(x,Q^2)f_{i}$ is  the  scale-dependent  nuclear 
modifications encoded in nPDFs (Fig.~\ref{fig:pPb}, right).

\section{Event selection and signal extraction}
This analysis is restricted to events that were filtered online by requiring the presence of at least one muon (electron) candidate 
with transverse momentum (energy) $p_{\rm{T}} > 12\rm{GeV}$ ($E_{\rm{T}} > 20\rm{GeV}$). 
Events are required offline to contain exactly one muon or electron candidate, 
with $p_{\rm{T}} > 30\rm{GeV}$ and $|\eta| < 2.1$, excluding in the electron case the transition region $1.444 < |\eta| < 1.566$ 
between the barrel and endcap. Both muon and electron candidates are required to be isolated from nearby hadronic activity 
within a cone of $0.3$ around the direction of the track at the primary event vertex. 
Events are further required to have at least four reconstructed jets--using the anti-$k_{\rm{T}}$ clustering algorithm with 0.4 cone size--with $p_{\rm{T}}> 25\rm{GeV}$ and $|\eta|< 2.5$,
that are separated by at least $0.3$ from the selected muon or electron in the $\eta$--$\phi$ plane.
Jets from b quarks are tagged based on the presence of a secondary vertex from B-hadron decays, identified using
a multivariate algorithm combining tracking information that corresponds to a b tagging efficiency of approximately 70$\%$
with a misidentification rate of less than 0.1$\%$ for light-flavor jets.
\begin{figure}[htb]
\centering
\includegraphics[scale=0.25]{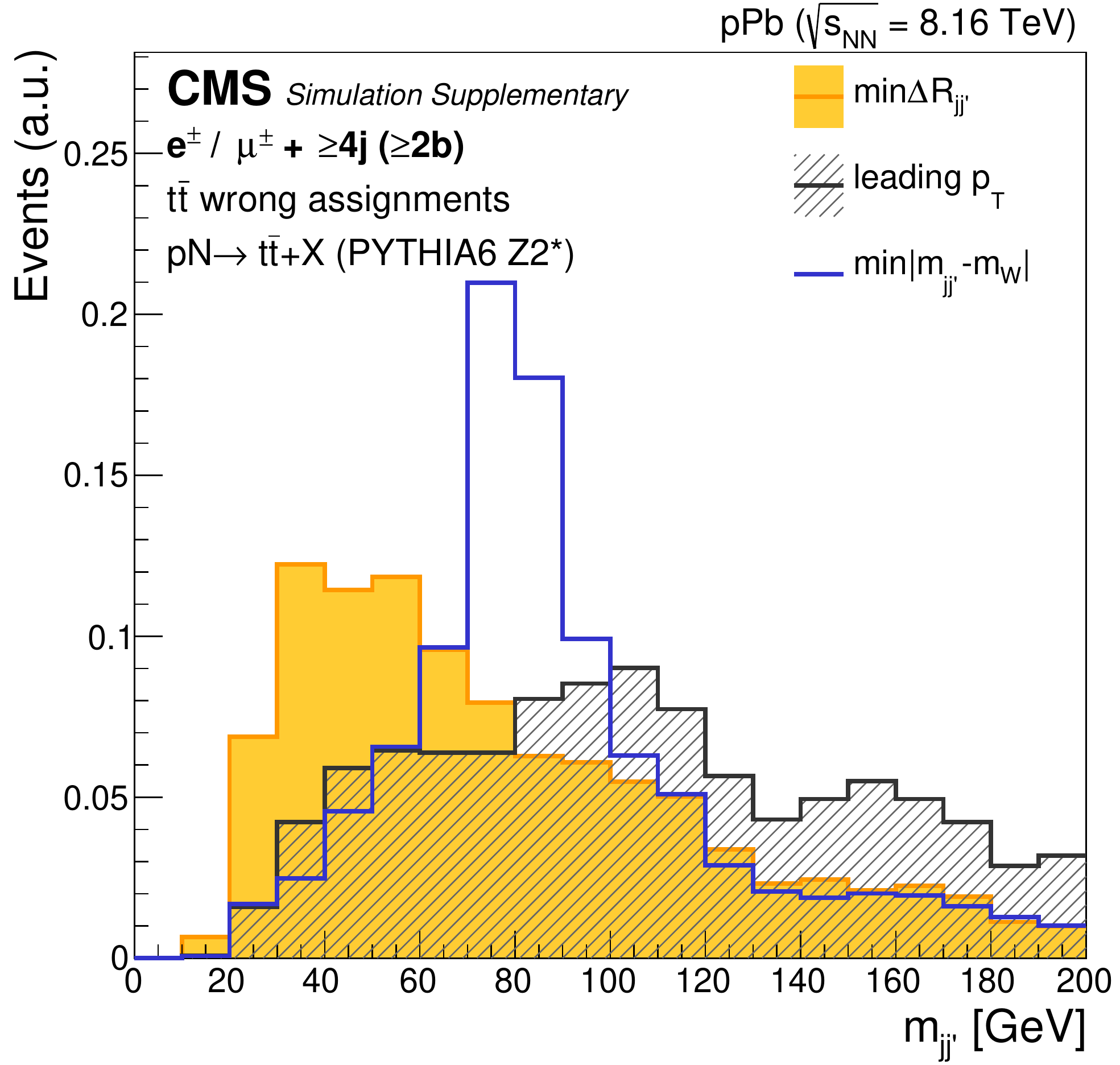}
\includegraphics[scale=0.245]{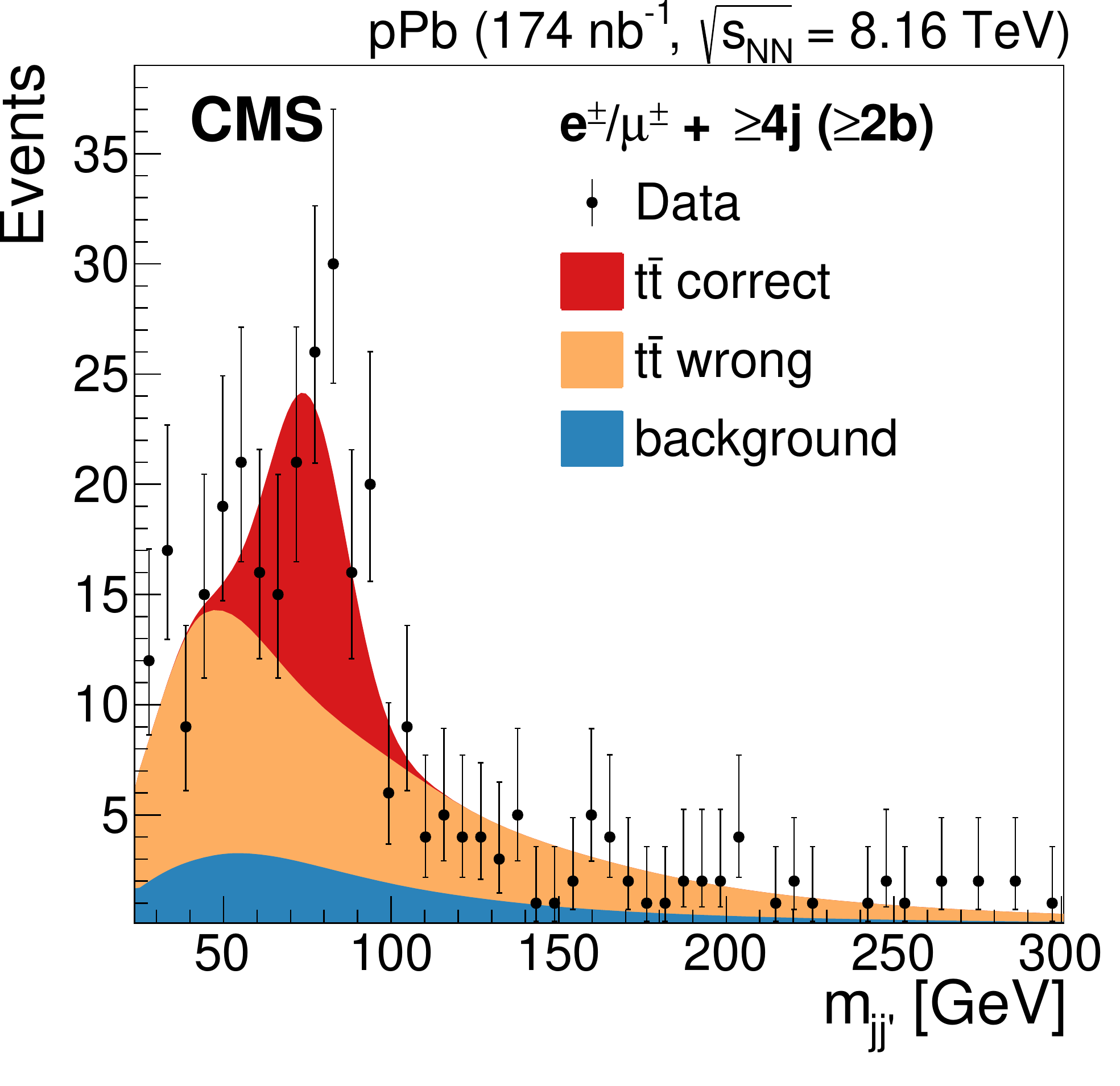}
\caption{
(left) Dijet invariant mass spectrum obtained by
using different algorithms in the pairing of the $\mathrm{j} \mathrm{j}^{\prime}$ system in the $\geq$2 b-tagged jet category; 
pairs where at least one reconstructed jet cannot be matched at parton level are shown~\cite{ttpPb}.
(right) Invariant mass distributions of the W candidate in the $\geq$2 b-tagged jet category after all selections.
The red and orange areas correspond to the signal simulation (correct and wrong assignments, respectively),
while the blue one corresponds to the estimated non-top background contributions~\cite{ttpPb}.
}
\label{fig:ressonant}
\end{figure}

The resonant nature of the invariant mass of $\mathrm{j} \mathrm{j}^{\prime}$, $m_{\mathrm{j} \mathrm{j}^{\prime}}$, 
provides a distinctive feature of the signal with respect to the main backgrounds, i.e., from QCD multijet and W+jets processes. 
In all cases, the $\mathrm{j} \mathrm{j}^{\prime}$ pair with the smallest separation in the $\eta$--$\phi$ plane is used to form a W 
boson candidate (Fig.~\ref{fig:ressonant}, left). The signal is separately modeled for correct (``pure'' resonant) and wrong assignments, respectively, while the background contribution 
from W+jets is assumed to be described by a Landau function, and that from QCD multijet is estimated with a nonparametric kernel approach (Fig.~\ref{fig:ressonant}, right).
The background parameterization is validated with the help of dedicated control regions.

\begin{figure}[htb]
\centering
\includegraphics[scale=0.25]{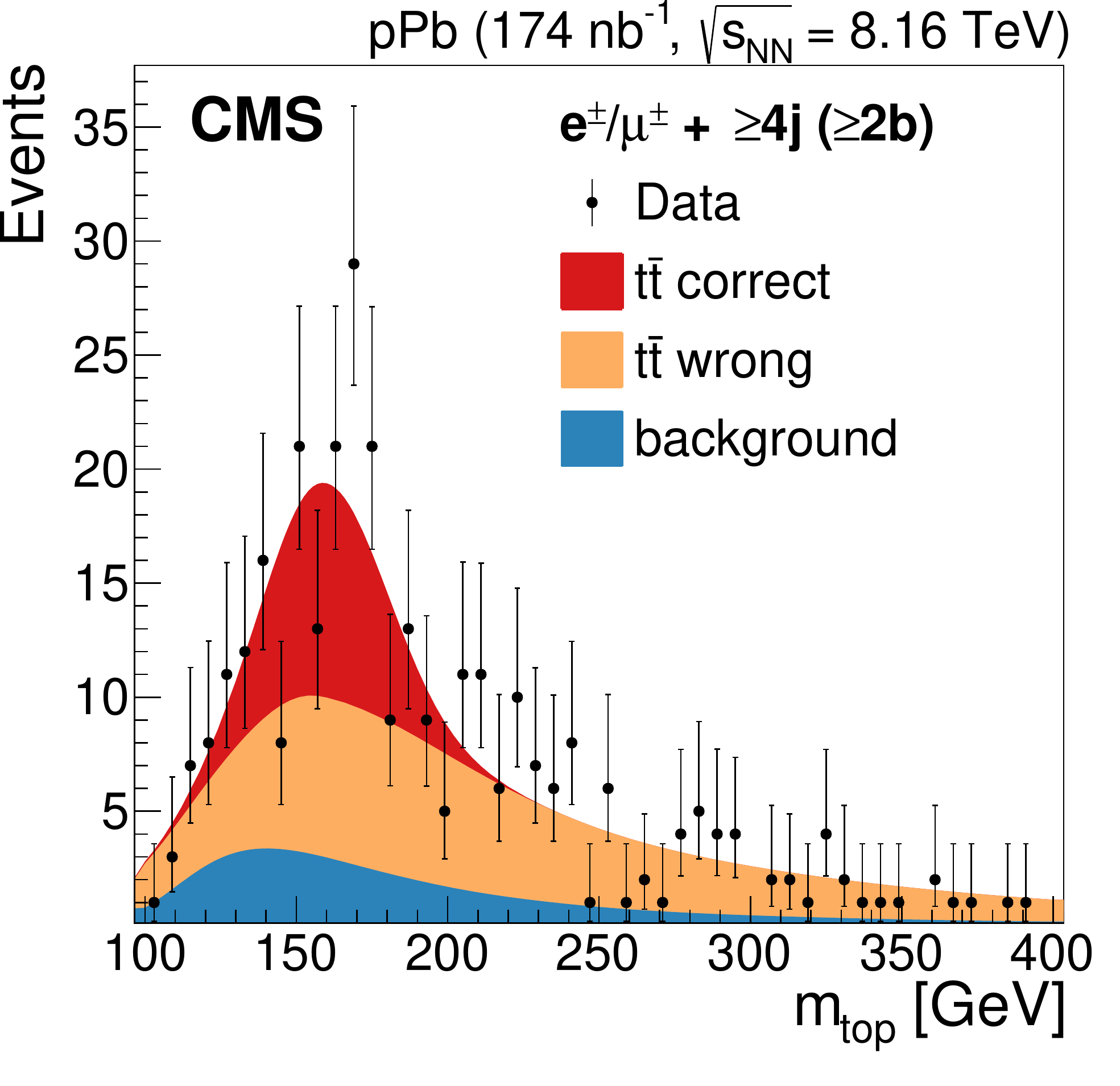}
\includegraphics[scale=0.35]{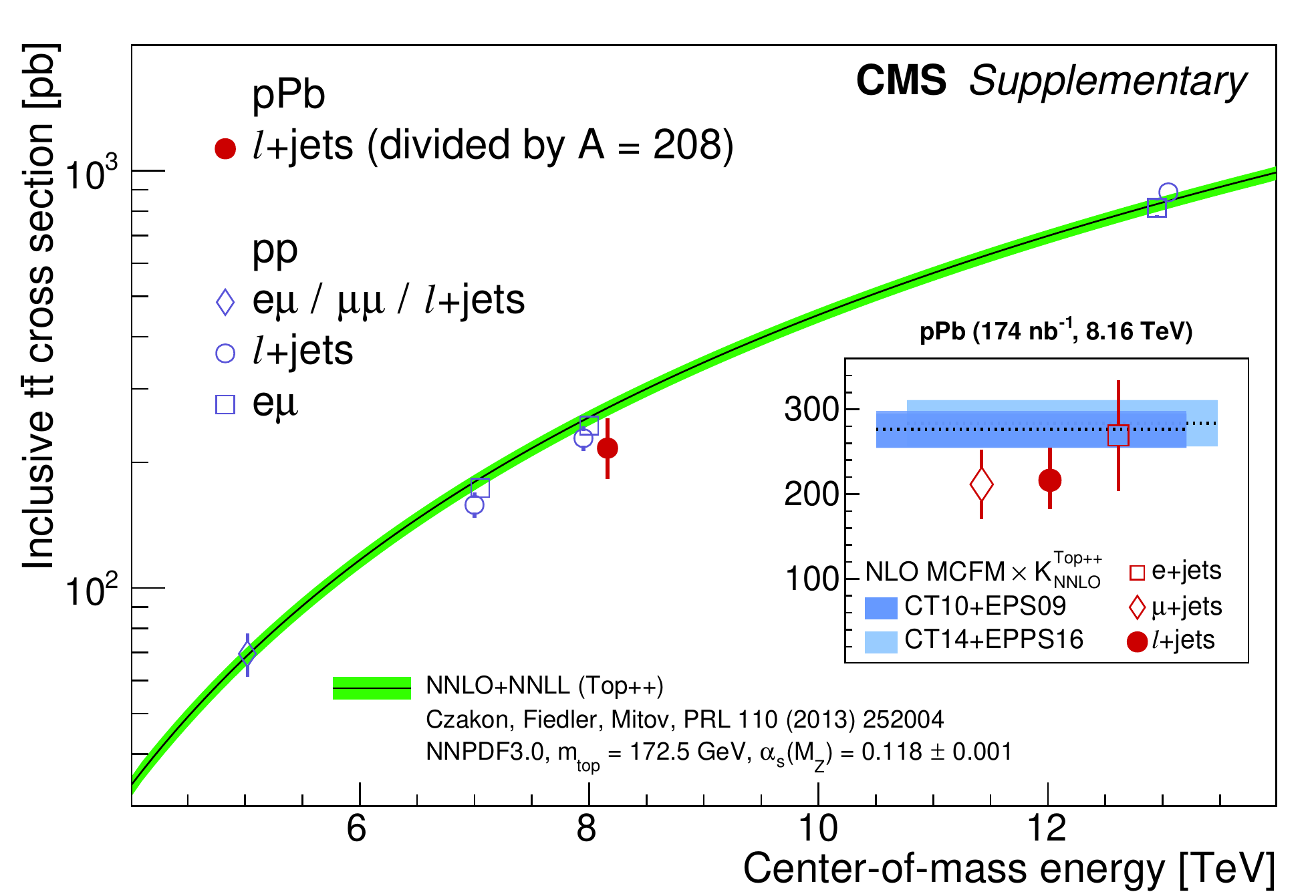}
\caption{
(left) Invariant mass distributions of the $\rm{t} \to \mathrm{j} \mathrm{j}^{\prime} b$ candidates
in the $\geq$2 b-tagged jet category after all selections.
All signal and background parameters are kept fixed to the outcome of the $m_{\mathrm{j} \mathrm{j}^{\prime}}$ fit~\cite{ttpPb}.
(right) Top quark pair production cross section in pp and pPb collisions collisions as a function of $\sqrt{s_{\rm{NN}}}$; 
the CMS measurements~\cite{ttpPb} are compared to the NNLO+NNLL theory predictions~\cite{alex} using state-of-the-art nPDFs~\cite{Eskola}.}
\label{fig:masses}
\end{figure}

The free parameters of the fit are the normalization of the signal, QCD multijet, and W+jets yields
(as well as the parameters of their functional forms), the b-finding efficiency, i.e.,
the probability that a jet originating from the b quark from a top quark decay passes both the kinematic 
and the b tagging selections, and an overall jet energy scale factor.
The combined fit to both channels and all three event categories yields a result of $45\pm8\ \rm{nb}$, 
with the total uncertainty having been obtained from the covariance matrix of the fit.
The statistical part of the total uncertainty amounts to $\sim5\%$, evaluated 
by leaving the cross section value to float in the fit and fixing all other parameters to their post-fit values.
Even with the most conservative assumptions, the background-only hypothesis is excluded with a significance above five standard deviations.
To further support the hypothesis that the selected data are consistent with the production of top quarks, 
a proxy of the top quark mass, $\rm{m_{top}}$, is constructed as the invariant mass of candidates formed by pairing the W candidate 
with a b-tagged jet; ambiguities are resolved by choosing the combination that results in the smaller value of 
the $|m_{jj'b}-m_{\ell\nu b}|$ difference (Fig.~\ref{fig:masses}, left).

\section{Summary}

The top pair production cross section has been measured for the very first time in 
proton-nucleus collisions, using pPb data at $\sqrt{s_{\rm{NN}}}=8.16\ \rm{TeV}$ with a total integrated luminosity of 174~nb$^{-1}$.
The measurement is performed by analyzing events with exactly one isolated lepton and at least four jets, 
and minimally relies on assumptions derived by simulating signal and background processes.
The significance of the $\rm{t}\bar{\rm{t}}$ signal against the background-only hypothesis is above five standard deviations.
The measured cross section is $45\pm 8$~nb, 
consistent with the expectations from scaled pp data as well as perturbative quantum chromodynamics calculations (Fig.~\ref{fig:masses}, right).
This first study clearly paves the way for further detailed investigations of top quark production in nuclear interactions (e.g.~\cite{liliana}),
providing in particular a new tool for studies of the hot and dense matter created in nucleus-nucleus collisions.

\end{document}

%% file: econfmacros.tex
\def\beq{\begin{equation}}
\def\eeq#1{\label{#1}\end{equation}}
\def\eeqn{\end{equation}}

\def\beqa{\begin{eqnarray}}
\def\eeqa#1{\label{#1}\end{eqnarray}}
\def\eeqan{\end{eqnarray}}

\let\bar=\overbar

\def\Dslash{\not{\hbox{\kern-4pt $D$}}}
\def\dslash{\not{\hbox{\kern-2pt $\del$}}}

\def\msb{{\bar{\ssstyle M \kern -1pt S}}}

